\documentclass[a4paper,11pt]{article}
\usepackage{pos}

\newcommand\Nf{N_{\mathrm{f}}}

\title{Update on $SU(2)$ with one adjoint Dirac flavor}

\author*[a]{Ed Bennett}
\author[b]{Andreas Athenodorou}
\author[c]{Georg Bergner}
\author[d,e]{Pietro Butti}
\author[a,f]{Biagio Lucini}

\affiliation[a]{Swansea Academy of Advanced Computing, Swansea University, Bay Campus, Swansea SA1 8EN, United Kingdom}
\affiliation[b]{Computation-based Science and Technology Research Center, The Cyprus Institute, 20 Kavafi Str., Nicosia 2121, Cyprus}
\affiliation[c]{University of Jena, Institute for Theoretical Physics, Max-Wien-Platz 1, D-07743 Jena, Germany}
\affiliation[d]{Instituto de Física Teórica UAM-CSIC, Nicolás Cabrera 13--15, Universidad Autónoma de Madrid, Cantoblanco, E-28049 Madrid, Spain}
\affiliation[e]{Departamento de Física Teórica, Universidad Autónoma de Madrid, Francisco Tomás y Valiente 7, Módulo 15, Cantoblanco, E-28049 Madrid, Spain}
\affiliation[f]{Department of Mathematics, Swansea University, Bay Campus, Swansea SA1 8EN, United Kingdom}

\emailAdd{e.j.bennett@swansea.ac.uk}

\abstract{We present an update of our ongoing study of the $SU(2)$ gauge theory with one flavor of Dirac fermion in the adjoint representation. Compared to our previous results we now have data at larger lattice volumes, smaller values of the fermion mass, and also larger values of $\beta$. We present data for the spectrum of mesons, baryons, glueballs, and the hybrid fermion-glue state, as well as new estimates of the mass anomalous dimension from both finite-size hyperscaling and the Dirac mode number, and discuss the implications of these data for the presence or otherwise of chiral symmetry breaking in this theory.}

\FullConference{%
The 39th International Symposium on Lattice Field Theory,\\
8th-13th August, 2022,\\
Rheinische Friedrich-Wilhelms-Universität Bonn, Bonn, Germany
}

\begin{document}
\maketitle

\section{Introduction}
Since the discovery of the Higgs boson at the Large Hadron Collider in 2012~\cite{ATLAS:2012yve,CMS:2012qbp},
there has been significant theoretical work to understand its nature;
is it a fundamental scalar
(with an unnatural~\cite{tHooft:1979rat} mass),
or is it a state that emerges from some physics beyond the Standard Model (BSM)?
Among the candidates being explored from the BSM side,
one avenue is that the scalar Higgs emerges
as a bound state of some strongly-interacting dynamical theory.
One route by which such a theory could give rise to a light scalar,
compatible with experimental observations,
is a theory that is nearly infra-red (IR) conformal,
and which has a large mass anomalous dimension.

\begin{figure}
  \includegraphics[width=\textwidth]{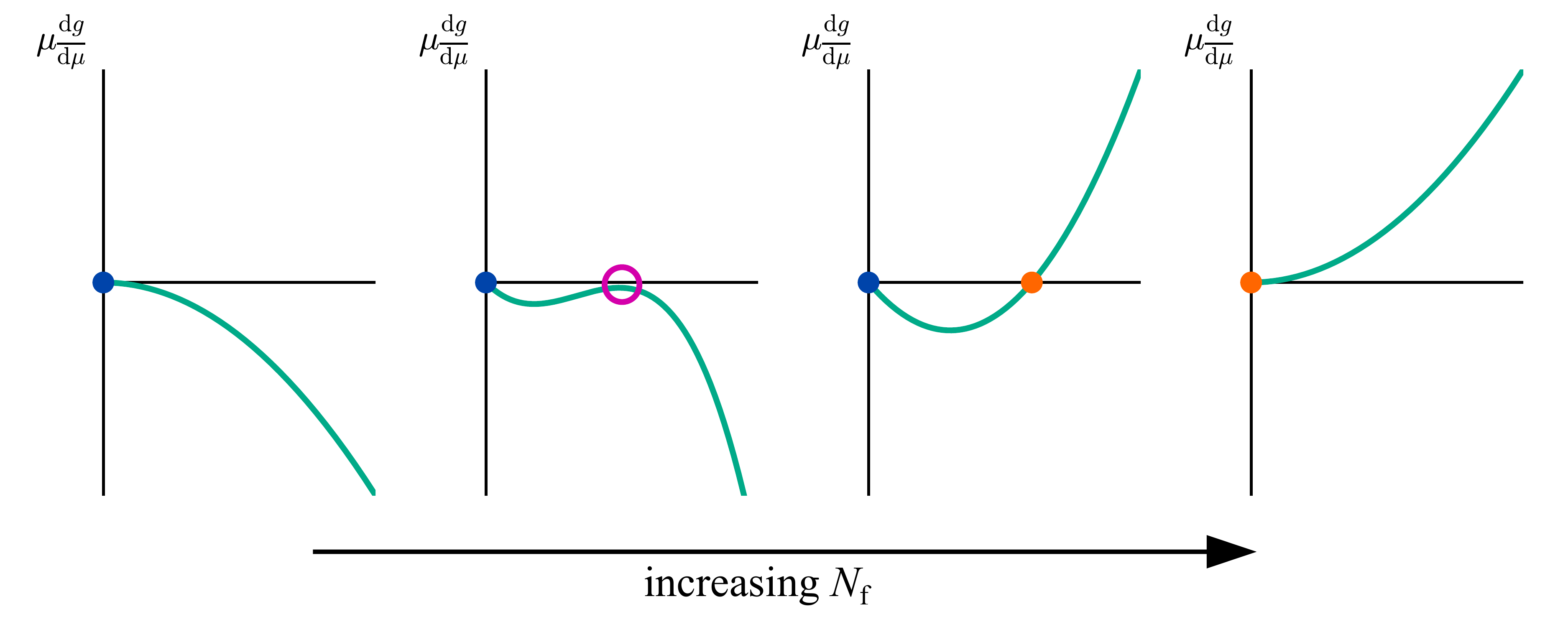}

  \caption{\label{fig:betafunctions}Cartoon of the possible shape of the beta functions
    of theories with various number of fermion flavours.
    On the left,
    at small number of flavours $\Nf$,
    the beta function decreases monotonically
    from the Gaussian ultraviolet (UV) fixed point marked in blue.
    This is the QCD-like case.
    On the right,
    at large number of flavours,
    it increases mononically from the Gaussian IR fixed point marked in orange.
    On centre-right,
    it initially decreases from the Gaussian UV fixed point in blue,
    but then increases again giving
    the Banks--Zaks fixed point marked in orange.
    This gives a theory that is conformal in the IR
    but still exhibits asymptotic freedom;
    this region of $\Nf$ is referred to as the \emph{conformal window}.
    Finally,
    at centre-left,
    where $\Nf$ is immediately below the lower edge of the conformal window,
    a would-be Banks--Zaks fixed point is approached but not reached before
    the beta function turns over again,
    marked in pink.
  }
\end{figure}

Non-Abelian gauge theories with number of fermion flavours $\Nf$
are expected to have a range of $\Nf$
(the \emph{conformal window})
for which
the theory has a conformal fixed point in the IR
while maintaining asymptotic freedom,
as illustrated in Fig.~\ref{fig:betafunctions}.
The exact extent of the conformal window
depends on the gauge group and the fermion representation.
While a perturbative estimate can be made~\cite{Dietrich:2006cm},
a first-principles lattice calculation is necessary
to verify whether or not this estimate is correct.

It is conjectured that immediately below
the lower edge of the conformal window,
there may be a region where a beta function
approaches a would-be fixed point,
and the theory exhibits a remnant of conformal behaviour.
Such theories are referred to as \emph{near-conformal};
they are anticipated to display long periods of slowly running coupling
(``walking''),
and a large value of the mass anomalous dimension.

The $SU(2)$ theory with one fermion flavour in the adjoint representation,
with the chiral symmetry breaking pattern $SU(2)\mapsto SO(2)\equiv U(1)$,
does not give sufficiently many Goldstone bosons to break electroweak symmetry,
and perturbative estimates~\cite{Dietrich:2006cm} place this theory well below
the lower end of the conformal window.
However,
studying this theory
provides input into knowing the location of the lower end of the conformal window,
and from there what additions may give
a theory that breaks electroweak symmetry
compatibly with experimental observations.

There is also interest in this theory from condensed matter theory,
as it may be dual
to the critical theory describing the evolution of
the phase transition between
a trivial insulator
and a topological insulator of the AIII class in $3+1$ dimensions~\cite{Bi:2018xvr}.

Our previous work~\cite{Athenodorou:2014eua,Athenodorou:2021wom}
has shown that the theory has
relatively constant ratios of bound state masses
in the region that has been able to be explored.
The scalar has been consistently observed to be the lightest state,
which is consistent with
other observations of conformal and near-conformal theories~\cite{Fodor:2016pls,DelDebbio:2015byq,LatKMI:2016xxi}.
The anomalous dimension,
measured both from finite size hyperscaling and from the Dirac mode number,
has been observed to be large,
but decreases as the value of $\beta$ considered increases.
Chiral perturbation theory meanwhile
has fitted the data increasingly poorly as $\beta$ increases.
Other work~\cite{Bi:2018xvr} has further found that
the hybrid glue--fermion state has
a mass compatible with that of the Dirac composite fermion.

Overall,
these findings are compatible with a theory
that is in or near the lower edge of the conformal window.
However, they do not confirm which of the two is the case.
They also have not yet been able to provide
a confirmed chiral or continuum limit extrapolation.

In this contribution,
we extend our previous study~\cite{Athenodorou:2021wom}
to three more values of $\beta$,
with the aim of being able to resolve one or more of these uncertainties.

\section{Lattice setup}

We make use of
the Wilson plaquette gauge action
and the Wilson fermion action without clover term.
We refer the interested reader to Ref.~\cite{Athenodorou:2014eua},
where we describe this setup in more detail.

We have generated ensembles at four values of $\beta$:
we have added two new ensembles at light masses to our previous set at $\beta=2.2$,
and also probed higher values of $\beta=2.25$, 2.3, and 2.4.
The region of masses of the state sourced by the $\gamma_{5}$ operator
(the $2^{+}$ scalar baryon)
is extended to
$am_{2_{\mathrm{s}}^{+}} \in (0.28, 1.11)$
($w_{0}m_{2{\mathrm{s}}^{+}}\in (1.0, 3.5)$).
We target the region $Lm_{2_{\mathrm{s}}^{+}}\gtrsim 10$
to control for finite-volume effects.

The lightest ensembles at $\beta=2.2$, 2.3, 2.4
are studied on a $L_{t}\times L_{s}=96\times48^{3}$ volume,
and the second-lightest ensembles on a $64\times32^{3}$ volume.
This allows exploration of
a region of significantly lower $am_{2_{\mathrm{s}}^{+}}$
than our previous work.
This was enabled by making use of Grid~\cite{Boyle:2015tjk,Gridrepo}
for the generation of these gauge ensembles,
as well as the new ensembles on a $48\times24^{3}$ volume,
allowing the use of GPU, rather than CPU resources.
We have continued to use HiRep~\cite{HiReprepo}
for generating ensembles at smaller volumes,
and for all observable measurements.

\section{Results}
\begin{figure}
  \center\includegraphics[width=\textwidth]{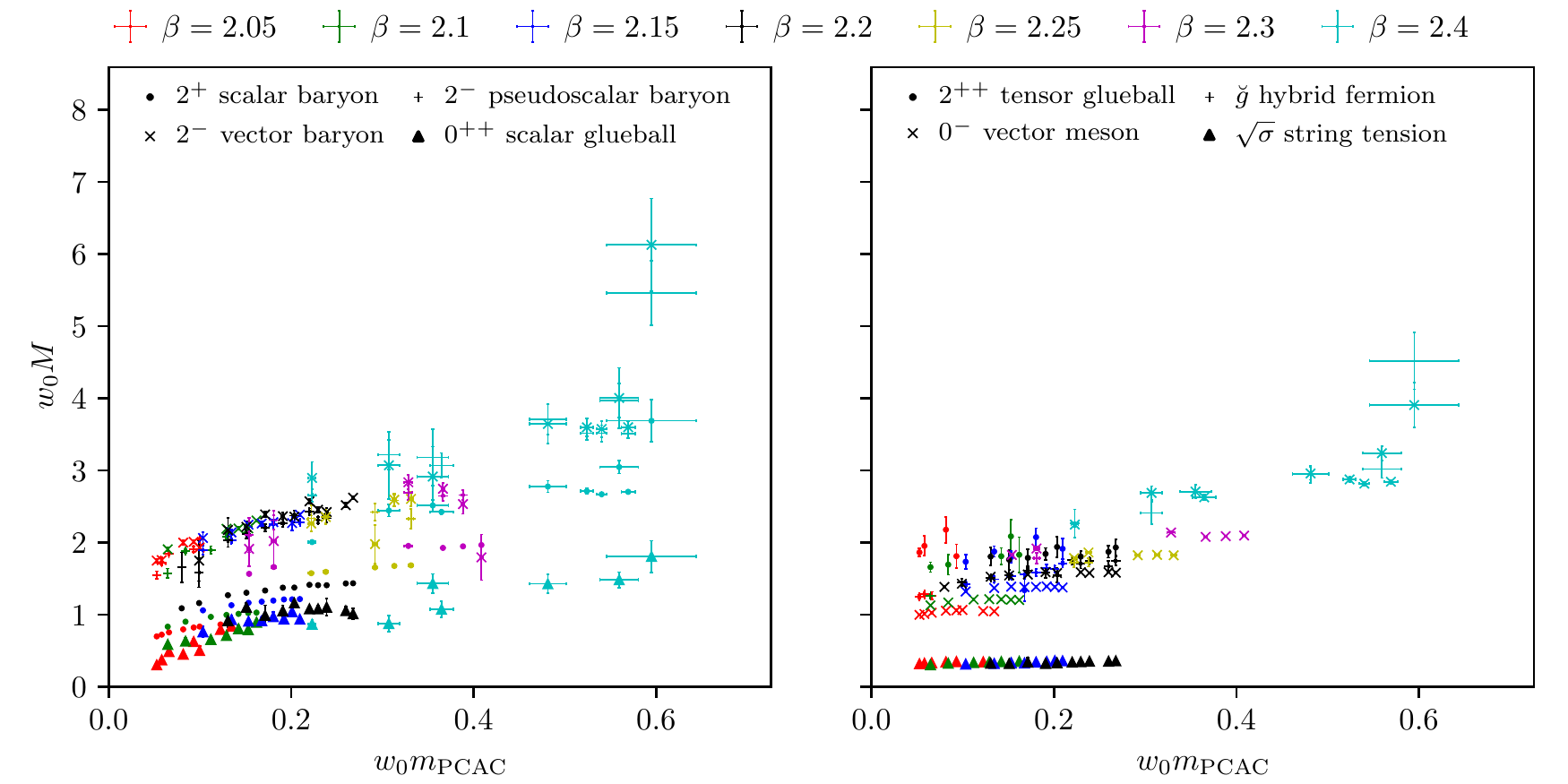}

  \caption{\label{fig:spectrum}(Preliminary.) Measured masses of the $2^{+}$ scalar, $2^{-}$ pseudoscalar, and $2^{-}$ vector baryon, the $0^{++}$ and $2^{++}$ glueballs, the $0^{-}$ vector meson, and the hybrid fermion-glue state $\breve{g}$, and the string tension $\sqrt{\sigma}$, as a function of the PCAC mass. All masses are normalised by the gradient flow scale $w_{0}$.}
\end{figure}

In Fig.~\ref{fig:spectrum}
we present preliminary results for the mass spectrum of the theory.
Some combinations of observable and ensemble
were still in the process of being computed at the time the contribution was presented;
these will be reported in a forthcoming journal publication~\cite{Athenodorou:2023tbc},
and the full finalised raw data and analysis workflow used to generate all plots
will be published concurrently with this.
For the data that are available,
the trend seen in our previous work continues.
The new data are relatively flat within each value of $\beta$,
with higher values of $\beta$ having successively slightly higher values.

In a conformal theory,
by definition there is no scale.
Thus when a conformal theory is simulated on the lattice
with a finite fermion mass,
then this deforming mass provides the only scale in the theory.
In such a mass-deformed conformal theory,
then one would expect all spectral quantities to scale in the same way;
that is,
ratios of spectral quantities would be constant.
As such,
our current data are consistent with near-conformal behaviour.
However,
flat spectral ratios are also seen in the heavy-fermion limit;
by themselves they are not conclusive evidence of
a conformal or near-conformal theory.

\begin{figure}
  \center\includegraphics[width=0.5\textwidth]{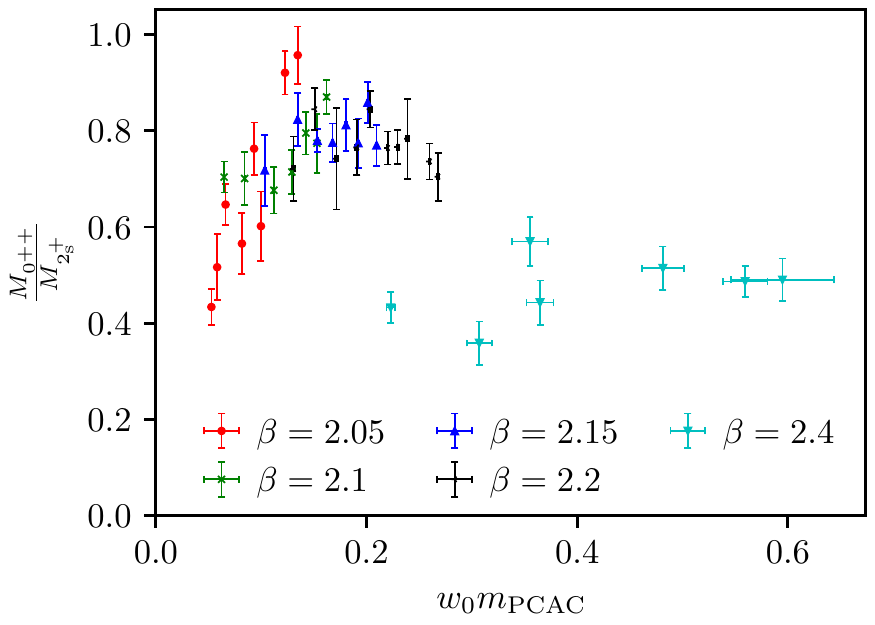}

  \caption{\label{fig:scalarratio}(Preliminary.) The ratio of the mass
    of the $0^{++}$ glueball to the $2^{+}$ scalar baryon,
    as a function of the PCAC mass in units of the gradient flow scale $w_{0}$.}
\end{figure}

As the adjoint representation of $SU(N)$ is real,
the spectrum of the theory includes
fermion-antifermion meson states
(computation of which includes connected and disconnected contributions)
and fermion-fermion baryon states
(which can be computed using only connected contributions).
In our presentation of the spectrum,
we adopt the convention we established in our earlier work~\cite{Athenodorou:2014eua},
labelling states by their quantum numbers under the unbroken $U(1)$ symmetry.

As such the state sourced by the connected $\gamma_{5}$ correlator,
which would be the $\pi$ meson in QCD,
we refer to here as the $2^{+}$ scalar baryon.
This is the would-be Nambu--Goldstone boson state;
in a theory with chiral symmetry breaking
it would be expected to be the lightest state in the spectrum.
Figure~\ref{fig:scalarratio} tests this
by plotting the ratio
of the mass of the scalar $0^{++}$ glueball
to that of the $2^{+}$ scalar baryon.
Previous work showed that the scalar glueball was the lighter state
throughout the region simulated.
The new ensemble at $\beta=2.4$ reinforces this,
showing a significantly lighter scalar than our previous ensembles.

\begin{figure}
  \center\includegraphics[width=\textwidth]{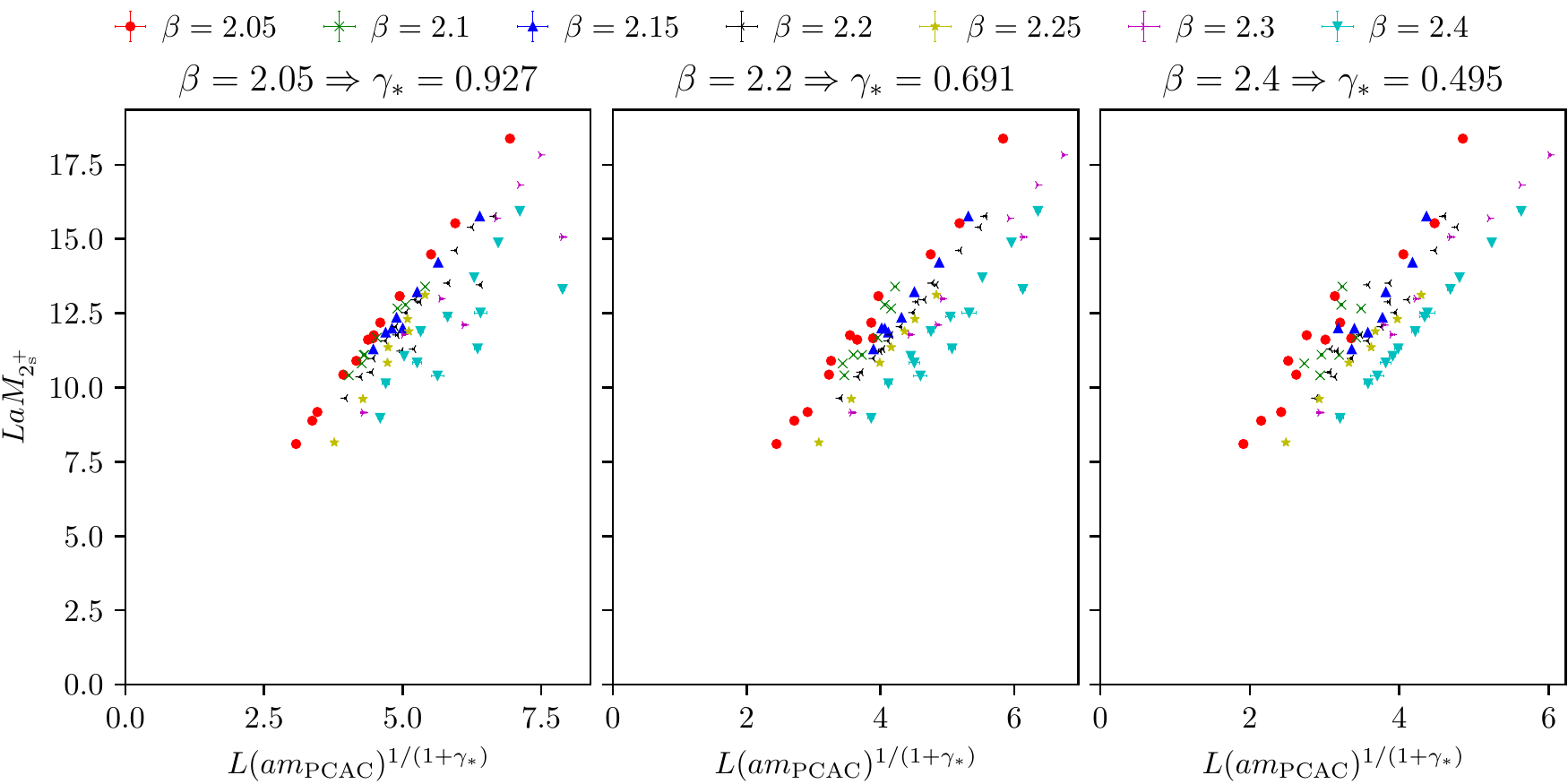}

  \caption{\label{fig:fshs}(Preliminary.) Curve-collapse fits
    of the $2^{+}$ scalar baryon spectrum
    to a finite-size hyperscaling Ansatz.
    Fits are shown for the three values $\beta=2.05$, 2.2, and 2.4.}
\end{figure}

State masses in a conformal (or near-conformal) theory
are expected to obey a finite-size hyperscaling relation of the form
\begin{equation}
  L\cdot am_{X}=f(L(m_{\mathrm{PCAC}})^{1/(1+\gamma_{*})})\;,
\end{equation}
where $m_{\mathrm{PCAC}}$ is the fermion mass from the PCAC relation,
and $\gamma_{*}$ is the mass anomalous dimension.
Fitting data from each value of $\beta$ studied to this functional form
using the method described in Ref.~\cite{Athenodorou:2021wom}
gives estimates for $\gamma_{*}$.
The values obtained continue to reduce as a function of $\beta$.
Three examples of these fits are shown in Fig.~\ref{fig:fshs};
in each case the data for one value of $\beta$ can be seen to line up visually.

\begin{figure}
  \center\includegraphics[width=\textwidth]{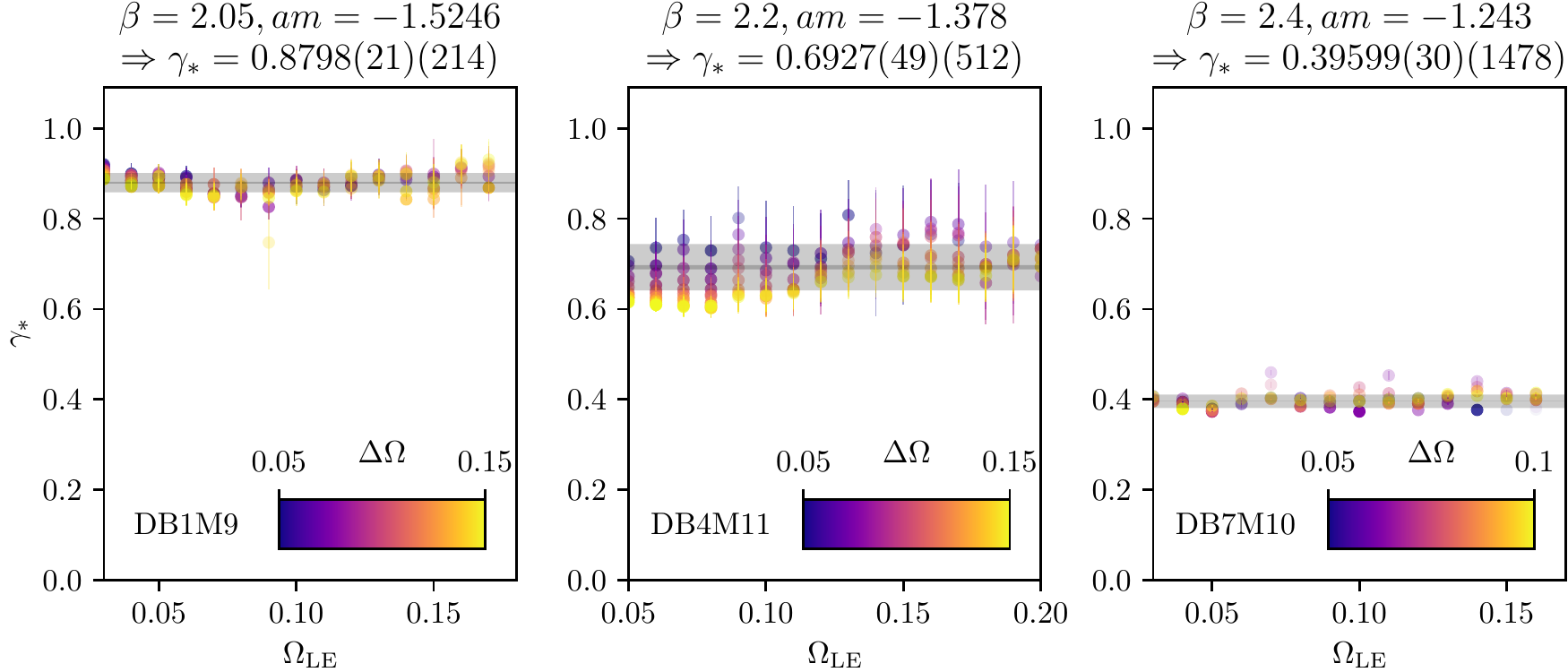}

  \caption{\label{fig:modenumber}(Preliminary.) Results of fits of the Dirac eigenmode spectrum
    to a conformal Ansatz for various windows of eigenvalue.
    The horizontal axis shows the lower end of this window,
    and the colour its length.
    The vertical axis shows the fitted anomalous dimension $\gamma_{*}$.
    The intensity of the colour indicates the weight used in the model average.
    The inner and outer shaded bands represent
    the statistical and systematic uncertainty
    on the estimate of $\gamma_{*}$ respectively.
  }
\end{figure}

An estimate for the anomalous dimension may also be obtained
by fitting the spectrum of eigenmodes of the Dirac operator.
We perform this analysis
in a similar manner to
the description in Ref.~\cite{Athenodorou:2014eua},
but use improved model averaging techniques to take the final fit.
Fuller details of these improvements will be discussed in
a forthcoming journal publication~\cite{Athenodorou:2023tbc}.
The results,
shown in Fig.~\ref{fig:modenumber},
show the same trend,
with the value obtained decreasing as the value of $\beta$ increases.

\begin{figure}
  \center\includegraphics[width=0.5\textwidth]{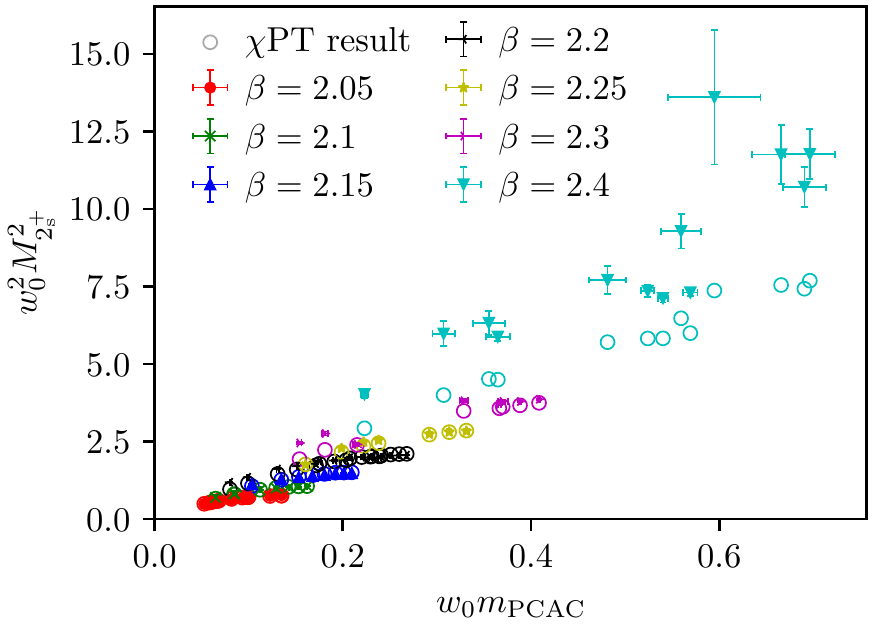}

  \caption{\label{fig:Xpt}(Preliminary.) Comparison of the data for the mass
    of the $2^{+}$ scalar baryon
    to a chiral perturbation theory fit of these data.
    Data points with error bars represent the measured data,
    while empty circles represent the fit result for that combination of $\beta$ and $m$.}
\end{figure}

In order to provide a fair comparison with
how well the conformal Ansatz matches observations,
we also fit our data for the $2^{+}$ scalar baryon
with an Ansatz informed by chiral perturbation theory:
\begin{align}
w_0^2 m_{2_{\mathsf{S}}^+}^2 =& \nonumber \; 2B\cdot w_0 m_{\mathrm{PCAC}}(1+LD_1 w_0 m_{\mathrm{PCAC}} \ln (D_2w_0 m_{\mathrm{PCAC}})) \\
  \nonumber &+W_1\cdot am_{\mathrm{PCAC}} \\
 &+ \frac{W_2}{w_0^2}\;,
\end{align}
where $B$, $D_{1}$, $D_2$, $W_{1}$, and $W_{2}$
are constants to be fitted,
and $w_{0}$ is a scale determined from the gradient flow~\cite{Athenodorou:2021wom}.

The result of this fit is shown in Fig.~\ref{fig:Xpt}.
We observe that as previously seen,
the quality of this fit continues to decrease as $\beta$ increases.

\section{Conclusions and outlook}

We have extended our study of $SU(2)$ with one adjoint Dirac flavour to
significantly lower fermion masses,
larger volumes,
and larger values of $\beta$
than we have previously studied.
The patterns we observed in our previous data continue to hold:
namely,
we continue to see near-flat spectral ratios,
the data continue to show finite-size hyperscaling
with an anomalous dimension $\gamma_{*}$ that decreases with $\beta$,
chiral perturbation theory continues to fit the data
increasingly poorly as $\beta$ increases,
and the scalar continues to be the lightest state observed.

The three possible conclusions stated in our previous work~\cite{Athenodorou:2021wom} remain valid.
The theory may be chirally broken and QCD-like,
with large lattice artefacts
and with simulations too far from the chiral limit.
The theory may be conformal with large scaling deviations.
Or the theory may be conformal,
with a stronger influence from the lattice bulk phase than anticipated.

Despite running significantly more computationally-intensive simulations,
the new information we have gained is relatively limited.
To extend the work to yet larger volumes or values of $\beta$,
or to yet lighter fermion masses,
while using the same lattice setup,
would require unreasonable amounts of computational resource,
and would give no guarantee of any new information.
As such,
our intention is to adjust our lattice setup for future work.
We plan to use a chiral lattice fermion action
to reduce the potential influence of lattice artefacts.
This is likely to be able to start eliminating
some of these possible interpretations.

\section{Acknowledgements}

The work of E.B. is supported by
the UKRI Science and Technology Facilities Council (STFC)
Research Software Engineering Fellowship EP/V052489/1.
A.A. has been financially supported by
the European Union’s Horizon 2020 research and innovation programme
``Tips in SCQFT''
under the Marie Skłodowska-Curie grant agreement No. 791122
as well as by the Horizon 2020 European research infrastructures programme ``NI4OS-Europe''
with grant agreement No.~857645.
G.B. is funded by the Deutsche Forschungsgemeinschaft (DFG)
under Grant No.~432299911 and 431842497.
P.B. acknowledges the financial support of
the European Union’s Horizon 2020 research and innovation programme
under the Marie Skłodowska-Curie grant agreement No.~813942.
The work of B.L. has been supported in part by
the STFC Consolidated Grants No. ST/P00055X/1 and No. ST/T000813/1,
and in part by the Royal Society Wolfson Research Merit Award WM170010
and by the Leverhulme Trust Research Fellowship No. RF-2020-4619.
B.L. further received funding from the European Research Council (ERC)
under the European Union’s Horizon 2020 research and innovation program
under Grant Agreement No. 813942.
Numerical simulations have been performed on the Swansea SUNBIRD cluster
(part of the Supercomputing Wales project)
and AccelerateAI A100 GPU system,
and on the DiRAC Extreme Scaling service at the University of Edinburgh.
Supercomputing Wales and AccelerateAI are part funded by
the European Regional Development Fund (ERDF) via Welsh Government.
The DiRAC Extreme Scaling service is operated by
the Edinburgh Parallel Computing Centre
on behalf of the STFC DiRAC HPC Facility (www.dirac.ac.uk).
This equipment was funded by BEIS capital funding
via STFC capital grant ST/R00238X/1 and STFC DiRAC Operations grant ST/R001006/1.
DiRAC is part of the National e-Infrastructure.

\bibliography{proceeding} \bibliographystyle{apsrev}

\end{document}